\documentclass[reprint,twocolumn,superscriptaddress,showpacs,nofootinbib,notitlepage,preprintnumbers,aps,prd,floatfix]{revtex4-2}
\usepackage[utf8]{inputenc}
\usepackage{graphicx}
\usepackage{latexsym,amsmath,amssymb,lmodern,float,url,IEEEtrantools}
\usepackage{natbib}
\usepackage{color}
\usepackage{microtype}
\usepackage{bbold}
\usepackage{slashed}
\usepackage{multirow}
\usepackage{xcolor}
\usepackage{mathrsfs}  
\usepackage{adjustbox}
\usepackage{physics}

\usepackage[colorlinks=true,backref=false, linktocpage=true,
citecolor=blue,urlcolor=blue,linkcolor=blue,pdfpagemode=UseOutlines]{hyperref}
\usepackage{cleveref}

\hypersetup{%
 bookmarksnumbered=true,
 pdftitle = {},
 pdfsubject = {},
 pdfauthor = {},
 pdfkeywords = {}
}

\usepackage{orcidlink}

\newcommand{\fnal}{\affiliation{Fermi National Accelerator Laboratory, 
Batavia, IL 60510, USA}}

\newcommand{\sqms}{\affiliation{Superconducting and Quantum Materials System Center (SQMS), Batavia, Illinois, 60510, USA.}}

\begin{document}
\preprint{FERMILAB-PUB-26-0133-SQMS-T}

\author{Henry Lamm\,\orcidlink{0000-0003-3033-0791}}
\email{hlamm@fnal.gov}
\fnal
\sqms

\title{No Quantum Utility from Hadron Masses?\\
No, Quantum Utility from Hadron Masses!}
\begin{abstract}
Is there a quantum utility in establishing the masses of hadrons? Here we show that the best response is that of Farr\'es, Cap\'o, and Davis: perhaps, perhaps, perhaps. This is interesting given the general case that particle physics demands quantum computers.
For stable hadrons, classical LQCD has achieved sub-percent
precision with no sign problem, and quantum computers
offer no advantage. For resonances, the Maiani-Testa theorem is an obstruction that quantum simulation is immune to. For nuclei, Wick contractions and
signal-to-noise are genuine
classical barriers.
Underlying these cases is a unified picture connecting the sign
problem to Wigner negativity and T gate cost.
This manuscript was drafted from extensive interaction with \textsc{Claude}.
\end{abstract}

\maketitle

A spectre is haunting quantum computing—the spectre of spectroscopy~\cite{Hardy:2024ric}. While it is well accepted that particle physics demands quantum computers~\cite{Jordan:2017lea,Troyer:2004ge}, a boundary haze persists with respect to the their necessity for specific problems such as hadron masses. An asymptotic quantum speedup is cold comfort if the crossover occurs at excessive system sizes. Beyond the asymptotic concerns, two more pitfalls threaten claimed quantum utility. First, when classical algorithm are granted the same sampling access to input data, the speedup may evaporate entirely~\cite{Tang:2018lmw}. Second, if the quantum circuit's non-Clifford content is low, the Gottesman-Knill theorem licenses efficient classical simulation~\cite{Gottesman:1997zz}, meaning that an algorithm which appears quantum may be classically simulable without approximation.

Before surveying where quantum utility may lie, we define what we mean by an \textit{answer}: we demand a rigorous, systematically improvable approximation to a non-perturbative QFT, meaning a calculation that converges to the continuum result as the UV cutoff $a^{-1}\to\infty$ and IR cutoff $L\to\infty$ are removed, with all sources of error, e.g. truncation, discretization, finite volume, and statistical, quantifiable and controllable in principle, not merely estimated by comparison with experiment.  

Euclidean Lattice field theory (LFT) is the only systematically improvable first-principles framework, resting on the rigorous Osterwalder-Schrader theorems~\cite{Osterwalder:1973dx,Osterwalder:1974tc,Luscher:1976ms}. Alternative formulations exist. Most notable is light-front quantization which is conceptually attractive but lack a systematic regulators that sufficiently preserves symmetries~\cite{Brodsky:1997de,Lamm:2016djr,Mondal:2020dkb,Li:2017mlw,Vary:2025yqo}.
Other approaches carry their own uncontrolled approximation: Dyson-Schwinger equations require a truncation of the infinite tower of Green's functions with no small expansion parameter~\cite{Roberts:1994dr,Maris:2003vk}; the functional renormalization group rely on the derivative expansion whose convergence is not guaranteed~\cite{Berges:2000ew,Dupuis:2020fhh}; and chiral perturbation theory, while systematically improvable order by order in $p/\Lambda_\chi$, is limited to energies well below $\Lambda_\chi \sim 1\,\text{GeV}$~\cite{Weinberg:1978kz,Gasser:1983yg,Scherer:2002tk}. Quantum simulation in the Kogut-Susskind Hamiltonian formulation inherits the systematic improvability of LFT --- with a rigorous UV cutoff $a^{-1}$ and IR cutoff $L^{-1}$, controllable discretization and volume errors, and a well-defined continuum limit. Thus, we confine ourselves to LFT. 

The flippant answer to when quantum computers outperform classical
ones in LFT is: real-time dynamics and finite density.
But this answer, while correct, obscures the nature
of the obstruction and risks understating the difficulty.
The common thread is the \textit{sign problem}: a pathology that arises
when the weight function in a path integral or Monte Carlo estimator
lack positive-definiteness, preventing efficient importance sampling.
In Euclidean LFT, the Boltzmann weight $e^{-S_E[\phi]}$
is typically real and positive, enabling efficient Monte Carlo sampling.
Sign problems arise when positivity is lost: at finite baryon
chemical potential $\mu_B$ the fermion determinant acquires a complex
phase $e^{i\theta}$, and the average sign
$\langle\sigma\rangle=\langle e^{i\theta}\rangle$ decreases with the
space-time volume $V$ as $e^{-V\Delta f}$, where $\Delta f$ is the free
energy difference between the full and phase-quenched
theories~\cite{Nagata:2021ugx}; for the $\theta$-term, the topological
charge weight $e^{i\theta Q}$ similarly destroys
positivity~\cite{Bazavov:2024spg}; and for real-time evolution, the
Minkowski weight $e^{iS_M[\phi]}$ is a pure phase,
making importance sampling impossible in principle~\cite{Alexandru:2020wrj,Hangleiter:2019yck}.
For context, the best classical LQCD methods probe the equation of state to $\mu_B/T \lesssim
3.5$ on lattices up to $64^3\times 16$~\cite{Borsanyi:2021sxv}.
The physically compelling regime---neutron-star densities, the QCD critical point, and the
hadron-to-quark transition---remains entirely inaccessible to direct Euclidean
simulation~\cite{deForcrand:2009zkb,Nagata:2021ugx}.

Quantum simulation evades the problem at its root: unitary time
evolution is sign-problem-free, and the Hilbert space
can be restricted to fixed baryon number.  This LFT intuition can be further bolstered by quantum information ideas to disperse the boundary haze.

The Gottesman-Knill theorem establishes that any quantum circuit
composed entirely of Clifford gates can be efficiently simulated
classically~\cite{Gottesman:1997zz,Aaronson:2004xuh}; an analogous
result holds for continuous-variable circuits built from Gaussian
operations, measurements, and input
states~\cite{Mari:2011zz,Mari:2012ypq}.
In both cases the boundary is set by Wigner negativity: injecting
magic gates generates negativity and lifts a circuit
beyond efficient classical
simulation~\cite{Veitch:2012ttw,Mari:2012ypq}, with the cost scaling
as $\mathcal{O}(2^t)$ for $t$ T~gates in the worst
case~\cite{Bravyi:2016yhm,Bravyi:2015mry} and refined bounds
available~\cite{Pashayan:2015cos,Bu:2019qed,Qassim:2021mcx}.
These results provide the theoretical underpinning for why the sign
problem is not merely a numerical inconvenience but a structural
feature: a theory whose path integral weight is sign-problem-free is
one whose quantum representation is non-negative.

The physical intuition of these computational results comes through the Wigner function. This function is a quasiprobability distribution over phase
space that provides a representation of a state $\rho$. For a LFT, the Wigner functional generalizes the finite-dimensional
Wigner function by promoting $(q,p)$ to fields
$\phi(x)$ and their conjugate momenta $\pi(x)$.
For a density matrix $\rho$, the Wigner functional for a gauge theory is
\begin{equation}
\label{eq:wigner}
  W_{ab}[\phi,\pi] = \int\!\mathcal{D}\chi\,
  \Bigl\langle\phi-\tfrac{1}{2}\chi\Big|\,
  U\,\rho\,\Big|\phi+\tfrac{1}{2}\chi\Bigr\rangle_{\!ab}
  e^{i\int d^3x\,\pi(x)\chi(x)},
\end{equation}
where $U$ is a Wilson line maintaining gauge covariance.
$W_{ab}$ is real-valued with non-negative marginals,
$\int\!\mathcal{D}\pi\,W_{ab}[\phi,\pi]=\langle\phi|\rho|\phi\rangle\geq 0$,
but is not necessarily non-negative.
Its matrix elements are built from the transfer matrix.
By Hudson's theorem~\cite{HUDSON1974249,Soto:1983avg}, a pure state has
$W\geq 0$ everywhere if and only if it is Gaussian~\cite{BrockWerner};
for mixed states this is not true and non-Gaussian states with
$W\geq 0$ exist~\cite{Mandilara:2009iwk,Albarelli:2018ujl} such as LFT ensembles.

Under Wick rotation $t\to -i\tau$, the Fourier kernel in
Eq.~(\ref{eq:wigner}) becomes a real exponential, so the Euclidean
analogue of $W$ is defined via a Laplace transform. This is generically unbounded and its marginals lose
their probabilistic interpretation.
This is fine for LFT because there one never computes $W_E$, but instead extracts the spectrum from the large-$\tau$ decay of correlators.
The unboundedness of $W_E$ becomes a obstruction only when
one attempts to recover real-time
information since the
inversion is ill-posed and requires regularization.

In Minkowski space, negativity of the Wigner function encodes the fact that the path integral cannot be described by any classical probability distribution
on phase space. The analog of negativity in Euclidean space is the sign problem where the probability ceases to be non-negative.  In these situations, the average sign is
\begin{equation}
  \langle\sigma\rangle
  = \frac{\int \mathcal{D}\varphi\; |w(\varphi)|\, e^{i\theta(\varphi)}}
         {\int \mathcal{D}\varphi\; |w(\varphi)|}
  \;=\; e^{-V \Delta f}.
\end{equation}
The connection to Wigner negativity can be understood via the resource theory of
non-classicality~\cite{Emerson:2013zse,Mari:2012ypq,Pashayan:2015cos}. For a state $\rho$ with $W_\rho$ over a phase space
$\Lambda$, define the negativity as the $L_1$ norm of the $W_\rho$
\begin{equation}
  M_\rho = \|W_\rho\|_1 = \sum_{\lambda \in \Lambda} |W_\rho(\lambda)|.
  \label{eq:negativity}
\end{equation}
Since $\sum_\lambda W_\rho(\lambda) = 1$, one has $M_\rho \geq 1$ if and only if $W_\rho \geq 0$ everywhere. $\langle\sigma\rangle$ can be defined with respect to the ``phase-quenched'' $|W_\rho|$ by
\begin{equation}
  \langle \sigma \rangle
  = \frac{\sum_\lambda W_\rho(\lambda)}{\sum_\lambda |W_\rho(\lambda)|}= \frac{1}{M_\rho}.
  \label{eq:sign_negativity}
\end{equation}
Thus $W_\rho \geq 0$ everywhere means $\langle\sigma\rangle = 1$ and there is no sign problem, but when $W_\rho$ takes negative values $\langle\sigma\rangle < 1$ and a sign problem arises.

These ideas can be extended from states to path integrals. To start, Pashayan, Wallman \& Bartlett~\cite{Pashayan:2015cos} extend this to quantum circuits. For a circuit with input state $\rho$, unitaries $U_1,\ldots,U_L$, and measurement effect $E$, they define per-element negativities
\begin{align}
  M_{U_l}(\lambda) = \sum_{\lambda'} |W_{U_l}(\lambda'|\lambda)|, \qquad
  \end{align}
and the maximum negativity $M_{U_l} = \max_\lambda M_{U_l}(\lambda)$.  Then the one can establish a negativity bound
\begin{equation}
  M_\to = \max_{\lambda_L} |W(E|\lambda_L)|\left(\prod_{l=1}^{L} M_{U_l}\right)M_\rho.
  \label{eq:Mforward}
\end{equation}
From this, they show the number of classical samples required to estimate a probability to precision
$\varepsilon$ with failure probability $\delta$ is bounded by
\begin{equation}
  s(\varepsilon,\delta) = \frac{2}{\varepsilon^2} M_\to^2 \ln(2/\delta).
  \label{eq:samplebound}
\end{equation}
which has the $\epsilon^2$ dependence one anticipates from importance sampling, and $M_{\rightarrow}=\langle\sigma\rangle^{-1}$. Classical simulation could be efficient if $M_\to$ grows polynomially in the system size. But negativity compounds multiplicatively through the circuit: even if each gate is only slightly negative, $M_\to$ can grow exponentially in $L$.

Taking each $U_i$ as a trotterized time evolution operator, the connection between a quantum circuit and a path integral formulation becomes clear~\cite{Aaronson:2016guw,Huang:2021wsp} and thus the sign problem finds a physical interpretation. 

A circuit of Clifford gates and $t$ $T$~gates is equivalent to a stabilizer circuit on
$|T\rangle\!\langle T|^{\otimes t}$~\cite{PhysRevA.71.022316}, so hardness reduces
to non-stabilizerness.  Pashayan~et~al.~\cite{Pashayan:2015cos}
showed that Monte Carlo estimation of probabilities costs
$\mathcal{O}(\mathcal{N}(\rho)^2/\epsilon^2)$, where $\mathcal{N}(\rho)$ is
the minimum $L_1$ norm of any decomposition of $\rho$.
Restricting to pure stabilizer projectors, $\mathcal{N}$ becomes the
\emph{robustness of magic} $R$ which grows exponentially in $t$~\cite{Howard:2017maw}.  Seddon~et~al.~\cite{Seddon:2021rij}
introduced a strictly smaller measure $\Lambda \leq R$ that is multiplicative
on tensor products of single-qubit states and coincides with other
magic proxies, unifying stabilizer-rank and quasiprobability approaches.
For $t$ $T$~state, $\Lambda = 4^{0.2284\,t}$, yielding a
simulator running in $\mathcal{O}(4^{0.2284\,t})$, an improvement
over~\cite{Seddon:2021rij}. Thus shows that each $T$~gate multiplies the
classical simulation cost by a factor of ${\approx}\,1.38$.

The multiplicative compounding of Eq.~\eqref{eq:Mforward} is why the
LFT sign problem $\langle e^{i\theta} \rangle \sim e^{-V \Delta f}$
decays exponentially in $V$. Each spacetime cell contributes a small
phase, and these compound, in exact analogy with
the factors $M_{U_l}$ in Eq.~\eqref{eq:Mforward}. With this perspective, one must ask the question, can one understand or quantify the limits of ameliorating sign problems from a quantum information perspective?

The answer is yes. Solving the sign problem requires finding a stoquastic basis for Hamiltonians -- where all off-diagonal entries are non-positive -- and is
NP-hard~\cite{Hangleiter:2019yck}. Instead, lattice theorists try to find local basis rotations which reduce its severity e.g.~\cite{Alexandru:2020wrj,Lamm:2018ocw}. Currently, the practicality of such methods is an active area of research~\cite{Lawrence:2021izu,Lawrence:2024pjg,Boguslavski:2024yto,Kanwar:2023otc,Hisayoshi:2025adi}. In~\cite{Hangleiter:2019yck}, the authors focus on a non-stoquasticity measure $\nu_1(H) = D^{-1}\|H_{\neg}\|_{\ell_1}$ where $D$ is the Hilbert space dimension, $H_{\neg}$ is the non-stoquasticity part of $H$. Unlike the average sign, $\nu_1$ is efficiently computable for local
Hamiltonians.
They show that
$\langle\sigma\rangle \sim \exp(-c\,D\nu_1(H))$, so minimizing
$\nu_1$ roughly minimizes the sign problem.
However, this relation is not universal:
highly non-stoquastic Hamiltonians can  have no sign problem,
and nearly stoquastic Hamiltonians with $\langle\sigma\rangle\rightarrow 0$, showing
the correspondence is generic but not rigorous.
Their main theorem sharpens this result: the
\textsc{SignEasing} decision problem - deciding whether a local basis
transformation can reduce $\nu_1$ below a given threshold - is
NP-complete for 2-local Hamiltonians~\cite{Hangleiter:2019yck} by relating it to \textsc{MaxCut}. Thus ameliorating sign problems is as hard as combinatorial optimization. The implication for LFT is stark: no
efficient classical algorithm can find an optimal
sampling basis to reduce the sign problem in the worst
case, placing a lower bound on the
difficulty of classical approaches. Armed with this structural picture, we survey the observables where quantum utility is established or anticipated.

A first cluster of observables concerns the early universe. In these situations, extreme non-equilibrium QFT processes --- electroweak baryogenesis, sphaleron-mediated baryon-number violation, vacuum decay, Schwinger mechanisms, and inflationary particle production --- occur and require real-time dynamics. Instanton tunneling has been measured on a quantum annealer~\cite{Abel:2020qzm}; bubble wall collisions have been simulated with tensor networks at hardware-relevant scales~\cite{Milsted:2020jmf}; and quantum algorithms for sphaleron evolution and particle-bubble wall collisions exist~\cite{Huang:2025nlo,Carena:2024peb}. Quantum simulations of the Schwinger mechanism have
been developed~\cite{Hidalgo:2023wzr,Draper:2025nku}. This program extends to cosmological particle production in an expanding spacetime~\cite{Maceda:2024rrd} and to false-vacuum nucleation~\cite{Ng:2020pxk}. 

A second cluster of real-time observables addresses scattering: the original quantum-utility proposal of Jordan, Lee and
Preskill~\cite{Jordan:2011ci} targeted S-matrix elements in scalar field theory,
and subsequent work has extended this to scattering amplitudes via the
LSZ reduction formula~\cite{Li:2023kex}, bound-state scattering~\cite{Turco:2023rmx},
phase shifts from wave-packet time delays~\cite{Gustafson:2021imb}, and finite-volume
corrections to scattering observables~\cite{Burbano:2025pef}. Complementing these physics targets, programme-level studies have mapped the landscape of quantum algorithms applicable to collider physics~\cite{Bauer:2025nzf}.

A third cluster considers parton structure: quantum algorithms exist for parton distribution functions~\cite{Lamm:2019uyc,Chen:2025zeh},
deep-inelastic-scattering structure functions~\cite{Mueller:2019qqj},
light-front parton correlators~\cite{Echevarria:2020wct},
quasiparton distributions and quasifragmentation functions in lower-dimensional gauge
theories~\cite{Grieninger:2024cdl,Grieninger:2024axp},
parton fragmentation~\cite{Li:2024nod},
and energy-energy correlators accessible at colliders~\cite{Lee:2024jnt}.

Fourth, transport phenomena encode how macroscopic behavior emerges from microscopic quantum dynamics — the viscosity of the quark-gluon plasma, the approach to hydrodynamics from far-from-equilibrium initial conditions, and the modification of hard probes as they traverse dense matter. Quantum simulation provides direct access to transport coefficients in gauge theories~\cite{Cohen:2021imf,Turro:2024pxu}, emergent hydrodynamic modes on plaquette chains~\cite{Turro:2025sec}, and the in-medium jet broadening and entropy growth relevant to heavy-ion collisions~\cite{Barata:2022wim,Barata:2023clv}.

The final cluster is the finite-density and topological sector of QCD, where sign problems arising from chemical potentials and the $\theta$-term render classical Monte Carlo exponentially costly. First steps toward quantum simulation have been taken: the $1+1d$ QCD phase diagram has been mapped on quantum hardware~\cite{Than:2024zaj}, variational methods demonstrated at finite temperature and density~\cite{Xie:2022jgj}, and chiral properties investigated in the Schwinger model with a $\theta$-term~\cite{Bazavov:2024spg,Kharzeev:2020kgc,Ikeda:2024rzv}. Qubit regularizations embed topological terms offering a concrete route to simulate topological phases~\cite{Bhattacharya:2023wuz}. Chiral fermions present an independent obstacle: the Nielsen--Ninomiya theorem forbids their naive lattice discretization~\cite{Nielsen:1981hk} and solutions can induce sign problems. Hamiltonian analogues of the circumventing formulations are only beginning to be constructed, with recent work delivering Ginsparg-Wilson Hamiltonians with tunable chiral symmetry~\cite{Singh:2025sye}.

What system sizes could be required for these various physics targets?  Using the relevant physical scales in the problem and general algorithmic considerations, one can estimate $N$ in terms of logical qubits.  A few examples are tabulated in Tab.~\ref{tab:lattice_requirements}; the results range from $10^4$ to $10^{36}$. These physical arguments can be integrated in specific digitization and algorithms to obtain order-of-magnitude estimate for gate costs as well~\cite{Kan:2021xfc,Hariprakash:2023tla,Rhodes:2024zbr}.

\begin{table*}
\renewcommand{\arraystretch}{1.4}
\caption{Estimated lattice parameters for quantum simulations of physics targets.
Benchmarking values used include: $E = p = 10\,\text{TeV}$ yielding $v_g=p/E\approx1$
(LHC center-of-mass scale) and $m = 1\,\text{GeV}$ (proton mass). For phase shifts, $\Delta E$ is the desired energy resolution; where $E \sim 2m$ with
$\Delta E/E \sim 1\%$.
For a heavy quarkonium systems $r_B \sim (m\alpha_s)^{-1} \approx 0.2\,\text{fm}$. The hard
scale is $Q = E_{\rm jet} = 10\,\text{TeV}$ and
$\Lambda_{\rm QCD} \approx 200\,\text{MeV}$.
For TMDs we take $x_B = 10^{-3}$ (typical LHC small-$x$). $T_{RHIC}\approx 300$ MeV and the mean free path is
$\ell_{\rm mfp} \sim (\alpha_s^2 T)^{-1}\approx 5\,\text{fm}$ with $\alpha_s \approx 0.3$.
The RHIC fireball size is $L_{\rm med} \sim 5\,\text{fm}$ and transverse
extent is $\Lambda_{\rm QCD}^{-1}$, and the hard jet scale is $p_T \sim 5\,\text{GeV}$.
Topological susceptibility requires $a \ll \rho_{\rm inst} \approx 0.3\,\text{fm}$
to resolve instantons.}
\begin{tabular}{lcccc}
\hline\hline
Physics target & $a$ & $L$ & $N_{\rm sites}$ & Logical qubits \\
\hline
Sphaleron rate
    & $\sim(5T_{EW})^{-1}\approx 0.3\,\text{fm}$
    & $\sim 10\,T_{EW}^{-1}\approx 1.5\,\text{fm}$
    & $\sim 50^3 \sim 10^5$
    & $\sim 10^6$ \\
EW bubble nucleation
    & $\sim(5m_H)^{-1}\approx 0.3\,\text{fm}$
    & $\sim 10\text{--}100\,T_{EW}^{-1}\approx 1.5\text{--}15\,\text{fm}$
    & $\sim 10^7\text{--}10^8$
    & $\sim 10^9$ \\
Schwinger mechanism
    & $\sim(5m_e)^{-1}\approx 0.08\,\text{pm}$
    & $\sim\text{few}\times m_e^{-1}$
    & $\sim 10^3\text{--}10^4$
    & $\sim 10^4\text{--}10^5$ \\
Inflationary particle production
    & $\sim H^{-1}$
    & $\sim\text{few}\times H^{-1}$
    & $\sim 100^3$
    & $\sim 10^7$ \\
\hline
Elementary particle $S$-matrix 
    & $\lesssim E^{-1} \sim 0.02\,\text{fm}$
    & $\gtrsim \sigma_x \sim \text{few}\times m^{-1} \sim 0.7\,\text{fm}$
    & $\sim (E/m)^3 \sim 10^{12}$
    & $\sim 10^{36}$
    \\
Scattering amplitudes (LSZ)
    & $\lesssim p^{-1} \sim 0.02\,\text{fm}$
    & $\sim \text{few} \times p^{-1} \sim 0.1\,\text{fm}$
    & $\sim (p/m)^3 \sim 10^{12}$
    & $\sim 10^{12}$
    \\
Phase shifts (time delays)
    & $\lesssim p^{-1} \sim 0.02\,\text{fm}$
    & $\gtrsim v_g/\Delta E \sim \text{fm}\times(E/\Delta E)$
    & $\sim (p/\Delta E)^3 v_g^3$
    & $\sim 10^4$--$10^5$
    \\
Bound-state scattering
    & $\lesssim r_B/5 \sim 0.04\,\text{fm}$
    & $\gtrsim 5\,r_B \sim 1\,\text{fm}$
    & $\sim 10^3$--$10^4$
    & $\sim 10^4$--$10^5$\\
    \hline
PDFs
    & $\lesssim Q^{-1} \sim 0.002\,\text{fm}$
    & $\gtrsim \Lambda_{\rm QCD}^{-1} \sim 1\,\text{fm}$
    & $\sim 10^9$
    & $\sim 10^{10}$
    \\
DIS structure functions
    & $\lesssim Q^{-1} \sim 0.002\,\text{fm}$
    & $\gtrsim W^{-1} \sim (x_B m_p)^{-1}$
    & $\sim 10^9$
    & $\sim 10^{10}$
    \\
Light-front correlators
    & $\lesssim Q^{-1} \sim 0.002\,\text{fm}$
    & $L_\perp \gtrsim \Lambda_{\rm QCD}^{-1}$;\; $L_\parallel \gtrsim (x_B P^+)^{-1}$
    & $\sim 10^{17}$
    & $\sim 10^{18}$
    \\
Parton fragmentation
    & $\lesssim Q^{-1} \sim 0.002\,\text{fm}$
    & $\gtrsim \Lambda_{\rm QCD}^{-1} \sim 1\,\text{fm}$
    & $\sim 10^9$
    & $\sim 10^{10}$
    \\
Energy-energy correlators
    & $\lesssim E_{\rm jet}^{-1} \sim 0.002\,\text{fm}$
    & $\gtrsim \Lambda_{\rm QCD}^{-1} \sim 1\,\text{fm}$
    & $\sim 10^{12}$
    & $\sim 10^{13}$
    \\
    \hline
Transport coefficients 
    & $\lesssim (5T_{RHIC})^{-1} \approx 0.13\,\text{fm}$
    & $\gtrsim \ell_{\rm mfp} \sim \alpha_s^{-2} T_{RHIC}^{-1} \sim 5\,\text{fm}$
    & $\sim 40^3 \sim 6\times 10^4$
    & $\sim 10^6$
    \\
In-medium jet broadening
    & $\lesssim p_T^{-1} \sim 0.04\,\text{fm}$
    & $L_{\rm med} \sim 5\,\text{fm};\; L_\perp \sim \Lambda_{\rm QCD}^{-1}$
    & $\sim 10^6$
    & $\sim 10^7$--$10^8$
    \\
    \hline
Finite $\mu_B$ QCD  
    & $\lesssim (5\Lambda_{\rm QCD})^{-1} \approx 0.2\,\text{fm}$
    & $\gtrsim m_\pi^{-1} \approx 1.4\,\text{fm}$
    & $\sim 7^3 \sim 3\times 10^2$
    & $\sim 10^4$--$10^5$
    \\
Finite $\theta$ QCD 
    & $\lesssim 0.06\,\text{fm}\ll \rho_{\rm inst}$
    & $\gtrsim$ few $\times \rho_{\rm inst} \approx 2\,\text{fm}$
    & $\sim 30^3 \sim 3\times 10^4$
    & $\sim 10^5$--$10^6$
    \\
Chiral magnetic effect
    & $\lesssim (5T)^{-1} \approx 0.13\,\text{fm}$
    & $\gtrsim \ell_{\rm mfp} \sim (\alpha_s^2 T)^{-1} \sim 5\,\text{fm}$
    & $\sim 40^3 \sim 6\times 10^4$
    & $\sim 10^6$
    \\
    \hline
    $a_1(1260)\rightarrow\rho\pi\rightarrow\pi\pi\pi$ & $\lesssim m_{a_1}^{-1} \approx 0.1\,\text{fm}$
& $\gtrsim \text{few}\times m_\pi^{-1} \sim 5\text{--}10\,\text{fm}$
& $\sim 75^3 \sim 4\times10^5$
& $\sim 10^6$--$10^7$ \\
    $^{40}$Ar mass & $\lesssim (5\Lambda_{\rm QCD})^{-1} \approx 0.2\,\text{fm}$
& $\sim 2r_0 A^{1/3} \approx 8\,\text{fm}$
& $\sim 40^3 \sim 6\times10^4$
& $\sim 10^6$--$10^7$ \\
\hline\hline
\end{tabular}
\label{tab:lattice_requirements}
\end{table*}

These results establish that a buffet of observables lies within the scope of quantum computation. The astute reader will have noticed that hadron masses are absent. This omission was deliberate.
Whether quantum computers offer any utility here depends on what your definition of ``hadron masses'' is.  In common parlance, a hadron is a composite subatomic particle made of two or more quarks held together by the strong nuclear force.  The nuance arises with the word ``particle". One subtlety is that unstable particles, or resonances, appear as a pole in the complex energy plane of a scattering amplitude rather than as an asymptotic state.  The other subtlety comes from nuclei, which some would say are bound states of hadrons. Following USQCD~\cite{Detmold:2019ghl,Kronfeld:2019nfb}, we distinguish between hadrons, resonances, and nuclei. For each class we can discuss quantum utility.

For hadrons stable under the strong interaction, the answer is an unambiguous no. Classical LQCD has, in a very real sense, shaken off any quantum competition. As early as 1981, 
pioneering quenched calculation achieved $10\%$ precision~\cite{Hamber:1981zn,Weingarten:1982qe}, and forty years of algorithmic and
hardware progress have brought this to the sub-percent
level~\cite{Hu:2024mas}.
QED and isospin-breaking corrections, while omitted in the most precise
mass calculations, are not a fundamental obstacle~\cite{FlavourLatticeAveragingGroupFLAG:2024oxs}.
For stable hadrons the sign problem is absent, importance sampling
works, and classical Monte Carlo has no known competitor --- quantum
computers offer no obvious utility and face an enormous precision
deficit to overcome.

Hadronic resonances occupy a qualitatively harder regime than stable hadrons, owing to a
fundamental obstruction captured by the Maiani--Testa
theorem~\cite{Maiani:1990ca}: in infinite volume, the Euclidean kernel $e^{-\omega t}$
projects onto threshold kinematics as $t\to\infty$, rendering scattering amplitudes above
threshold inaccessible from two-point functions.
Because a resonance is not an asymptotic state it cannot be read off directly from a
correlator; instead one must extract a matrix of finite-volume energy levels and invert the
L\"uscher quantization conditions to recover the infinite-volume scattering amplitude and
pole position~\cite{Luscher:1990ux}.
This indirect route has driven substantial progress in lattice QCD
spectroscopy~\cite{Briceno:2017max}, including determinations of the $\rho$ meson
mass~\cite{Lang:2011mn,Feng:2010es}, yet it introduces compounding difficulties: expensive
multi-volume and multi-frame ensembles; model-dependent analytic continuation for
coupled-channel resonances such as the $\sigma/f_0(500)$ and exotic hadrons near threshold;
and rapidly deteriorating signal-to-noise for multi-baryon
states~\cite{Parisi:1983ae,Lepage:1989hd,Wagman:2016bam}.
Smeared spectral reconstruction offers only limited relief given its ill-posed inverse
problem~\cite{Bruno:2020kyl,Carrillo:2024chu}.
Quantum simulation sidesteps the obstruction entirely: working in Minkowski signature, the
LSZ reduction formula applies directly and resonance poles are readable from the time-domain
signal~\cite{Gustafson:2021imb,Turco:2023rmx,Burbano:2025pef}---the Maiani--Testa theorem
is a statement about the information content of Euclidean correlators, not a technical
inconvenience to be engineered around.

Whether quantum simulation will achieve practical utility for resonance physics before
competitive fault-tolerant resources exist remains an open question. Classical computing
will likely resolve at least some above-threshold resonances in the interim, though
genuinely complex multi-body decays - such as
$a_1(1260)\to\rho\pi\to\pi\pi\pi$ - may prove intractable without it. Realising the
promise of quantum simulation in this context will require substantial advances in
formalism and algorithms beyond what is currently available~\cite{Ciavarella:2020vqm};
notably, the L\"uscher method is unlikely to play a role~\cite{Hardy:2024ric}. For resonances, the verdict on quantum utility is
therefore a cautious \emph{perhaps}.

Classical LQCD faces two compounding obstacles for nuclei: an extreme version of SNR degradation for above-threshold states $\sim \exp(-(M_N - \frac{3}{2}m_\pi)At)$~\cite{Parisi:1983ae,Lepage:1989hd,Wagman:2016bam} and a combinatorial explosion in the Wick contraction requiring $\mathcal{O}((3A/2)!^2)$ contractions~\cite{Yamamoto:2022jnn} over the quark fields. The current classical frontier reaches only two- and three-baryon systems near the physical pion
mass~\cite{Detmold:2019ghl}, with larger nuclei out of reach.

Quantum simulation offers a path around both obstacles: prepare a state
with overlap with the nuclear ground state at fixed quantum numbers,
then measure its energy.
Three strategies exist with distinct resource profiles.
QPE applies controlled $e^{-i\hat{H}t}$ to a trial state and reads off
the energy from the ancilla phase at precision $\epsilon$ with circuit
depth $\mathcal{O}(\epsilon^{-1})$~\cite{Yamamoto:2022jnn,Dumitrescu:2018njn}.
VQE minimizes a parametrized ansatz variationally but is limited at
large $A$ by the barren plateau
problem~\cite{Dumitrescu:2018njn,Robin:2023pgi}.
Adiabatic preparation deforms continuously from a tractable initial
Hamiltonian to full QCD, requiring evolution slow relative to the
instantaneous gap $\Delta(\lambda)$~\cite{Yamamoto:2022jnn}.
All three approaches share a common challenge: the nuclear excitation
spectrum is dense, with collective rotational gaps
$\Delta_{\rm rot}\sim\hbar^2/(2\mathcal{I})\propto A^{-5/3}$
shrinking with system size, and level crossings such as the
liquid-gas transition cause the gap to close entirely, diverging the
adiabatic runtime as $\mathcal{O}(\Delta^{-2})$ and demanding longer
QPE coherence times and flatter VQE
landscapes~\cite{Yamamoto:2022jnn}.
Despite this, the quantum scaling away from crossings is
$\mathcal{O}(A^{16/3})$, polynomially better than the classical
$\mathcal{O}(e^{c_2 A}A^4)$~\cite{Yamamoto:2022jnn}, with the
crossover favoring quantum simulation at modest baryon number.

As a concrete illustration, consider $^{40}$Ar.
The nucleus requires a box $L\sim 8\,\text{fm}$, spacing
$a\lesssim 0.2\,\text{fm}$, giving $N_{\rm sites}\sim 6\times10^4$
sites and $\sim10^6$--$10^7$ logical qubits; precision $\epsilon\lesssim 1\,\text{MeV}$ is needed to resolve
nuclear structure, a relative precision of
$\epsilon/M(^{40}\text{Ar})\sim3\times10^{-5}$.
QPE requires $\mathcal{O}(\epsilon^{-1})$ steps each costing
$\mathcal{O}(N_{\rm sites})$ gates, with an additional
$\mathcal{O}(\Delta^{-2})$ overhead from adiabatic state preparation for the liquid-gas transition gap
$\Delta\sim\Lambda_{\rm QCD}/A\sim5\,\text{MeV}$, giving a total gate
count of $\sim10^{10}$--$10^{12}$.
The $\mathcal{O}(A^{16/3})$ nuclear-operation scaling of
Ref.~\cite{Yamamoto:2022jnn} gives $40^{16/3}\sim4\times10^8$,
better than the classical
$\mathcal{O}(e^{c_2 A}A^4)\sim2\times10^{11}$ at $c_2\approx0.3$;
the apparent discrepancy with the $10^{10}$--$10^{12}$ gate estimate
reflects the different levels of the algorithm stack being counted. The latter includes only nuclear operations vs. the former where the full circuit overhead is accounted for. Since both the quantum and classical costs carry the same overhead, the utility remains intact.

The central question posed by this paper admits a three-part answer.
For stable hadrons, the answer is an unambiguous no: classical LQCD has
achieved sub-percent precision with no sign problem,
and quantum computers offer no structural advantage.
For resonances, the answer is perhaps: the Maiani-Testa
theorem~\cite{Maiani:1990ca} establishes a fundamental obstruction to
Euclidean methods that quantum simulation is
immune to, but the requisite algorithms remain
immature~\cite{Burbano:2025pef,Ciavarella:2020vqm}.
For nuclei, the answer trends toward yes: factorial Wick contraction
costs and exponential SNR degradation represent
genuine classical barriers, while quantum
simulation scales polynomially~\cite{Yamamoto:2022jnn}.

Undergirding all three cases is a unified picture connecting the sign
problem to Wigner negativity: the NP-completeness of
\textsc{SignEasing}~\cite{Hangleiter:2019yck} places a rigorous lower bound on classical approaches to
QCD, while the Pashayan-Wallman-Bartlett
bound~\cite{Pashayan:2015cos} quantifies how negativity compounds multiplicatively through a circuit, in exact analogy with the
volume-exponential decay of the average sign.
The boundary between classical efficiency and quantum utility is not
merely a matter of prefactors, but has a structural character.

Looking forward, Table~\ref{tab:lattice_requirements} reveals a striking
hierarchy: near-term targets at $\sim 10^4$--$10^5$ logical qubits, like phase shifts and finite-$\mu_B$ QCD, are within reach of projected hardware, while LHC-scale
scattering remains out of reach.
The nuclear problem stands out as uniquely well-motivated, and
developing end-to-end resource estimates for nuclei
should be a near-term priority.
We close by noting that the standard of systematic improvability adopted here has direct implications for near-term demonstrations:
current noisy-hardware results in lower-dimensional toy models are
valuable proofs of principle, but do not constitute answers in the
sense defined.
The path to systematically improvable Minkowski-signature QCD is long,
but the case that it can yield new physics is on firm footing.

\begin{acknowledgments}
    This paper was written by \textsc{Claude} with the human author acting largely as supervisor and editor. Despite this, HL thanks D. Hackett,
    /G. Fleming, M. Wagman, and E. Gustafson for comments and insight during the drafting of this manuscript. This work was supported by the U.S. Department of Energy, Office of Science, Office of High Energy Physics Quantum Information Science Enabled Discovery (QuantISED) program ``Toward Lattice QCD on Quantum Computers." This work was produced by Fermi Forward Discovery Group, LLC under Contract No. 89243024CSC000002 with the U.S. Department of Energy, Office of Science, Office of High Energy Physics.
\end{acknowledgments}

\bibliography{wise.bib}
\end{document}